\documentclass[letterpaper, 10 pt, conference]{ieeeconf}  
                                                         \usepackage[utf8]{inputenc}
\usepackage[T1]{fontenc} 

\IEEEoverridecommandlockouts                              
\usepackage{graphicx,url}
\usepackage[utf8]{inputenc}  
\usepackage{amsmath}
\usepackage{comment}
\usepackage{url}
\usepackage{bbm}
\usepackage{subfig}
\usepackage{color}				
\usepackage{graphicx}	
\usepackage{epsfig}
\usepackage[colorinlistoftodos]{todonotes}
\usepackage{authblk}

\usepackage
	{todonotes}
 \usepackage{fancyhdr} 
\usepackage{kantlipsum} 
\fancypagestyle{plain}{ 
\fancyhf{} 
\fancyhead[C]{Conference on \LaTeX} 
 
} 
\usepackage{eso-pic}

\newcommand{\pubj}{_{ij}} 

\title{Biases in the Facebook News Feed: a Case Study on the Italian Elections}

\author[1]{Eduardo Hargreaves}
\author[2]{Claudio Agosti}
\author[1]{Daniel Menasché}
\author[3]{\\ Giovanni Neglia}
\author[4]{Alexandre Reiffers-Masson}
\author[3]{Eitan Altman}
\affil[1]{Federal University of Rio de Janeiro (Brazil), Email:eduardo@hargreaves.tech, sadoc@dcc.ufrj.br}
\affil[2]{University of Amsterdam (Netherlands), Email:claudio.agosti@tracking.exposed}
\affil[3]{Universit\'e C\^ote d'Azur, Inria (France), Email:\{giovanni.neglia, eitan.altman\}@inria.fr}
\affil[4]{Indian Institute of Science (India), Email:reiffers.alexandre@gmail.com}

\begin{document}

\maketitle
\IEEEoverridecommandlockouts

\begin{abstract}

 Facebook News Feed personalization  algorithm  has a significant impact, on a daily basis, on
 the lifestyle, mood and opinion of millions of Internet users. 
Nonetheless, the behavior
      of such  algorithms usually lacks transparency, 
      motivating measurements, modeling and
       analysis in order to understand 
       and improve its properties.
 %
 In this paper, we propose a reproducible  methodology encompassing measurements and an analytical model to capture the visibility of publishers over a News Feed. 
First,   measurements are used to parameterize and to validate the expressive power of the proposed model.  Then, we conduct a what-if analysis to assess the visibility  bias incurred by the users against  a baseline derived from the model.  Our results indicate that a significant bias exists and it is more prominent at the top position of the News Feed.  In addition, we found that the bias is non-negligible even for  users that are  deliberately set as neutral with respect to their political views. 
 %
\end{abstract}

\section{Introduction}





%
Online social networks (OSNs) are increasingly being used to share political opinion and news, particularly during election periods~\cite{jones2017social}.  %
%
With almost 2 billion  active users per month, Facebook is currently the largest OSN~\cite{Facebook_2bi}. 
As any platform for media sharing, one of the core functions of Facebook is to organize the available information to     improve user experience.  To that aim, Facebook counts on  a filtering and personalization product referred to as the \emph{News Feed} algorithm~\cite{Facebook_top_stories}.

In essence, the {News Feed} algorithm is a recommendation system that presents posts to users in their News Feed. However, unlike classical recommendation systems, e.g., for music and movies,  the {News Feed} poses its own set of challenges.   The  {News Feed} needs to account for content continuously  generated by multiple sources (publishers), and users typically do not provide explicit feedback about the recommendations.  In addition, the News Feed  algorithm may impact users that are usually unaware of its influence~\cite{Eslami2015}, e.g.,   by creating a \emph{filter bubble} that reinforces the perceived users' opinions~\cite{pariser2011filter}.
Clearly, the News Feed algorithm   influences the way political  discourse is perceived, potentially creating biases towards sources and ultimately impacting how democracy works.

In this paper we take important steps towards measuring, modeling and analyzing the bias caused by the News Feed algorithm.  We believe that a better understanding of the News Feed can be instrumental in the design of new filtering algorithms, which may eventually be actively controlled by users. Furthermore, obtaining this knowledge through public and reproducible  measurements is key for increasing the awareness of users about the filtering of the content they consume.

The literature on the News Feed algorithm is vast, including  measurements~\cite{Cheng2018,BESSI2016319,bessi2016users,Bucher2012}, models~\cite{altman2013competition,dhounchak2017viral} and  user awareness surveys~\cite{Eslami2015}.  
%
Nonetheless, most of the prior work that quantifies the effect of OSNs on information diffusion~\cite{Bakshy2012,Bakshy2015} relies on measurements obtained through restrictive non-disclosure agreements that are  not made publicly  available to other researchers and practitioners.   As the data analyzed in such studies is very sensitive, and their sources are not audited, there are multiple potential factors and confounding variables that are unreachable to the general public.  
Our goal is to provide insights on the filtering that occurs in OSNs through  a  reproducible methodology and a dataset that we have made publicly available~\cite{fbtrex}.


%
%

In this work, we propose a methodology that encompasses measurements and a model for the analysis of bias in OSNs.  Our key contributions are summarized as follows.

\textbf{A measurement methodology } is introduced to publicly and transparently audit the OSN ecosystem, focusing on the Facebook News Feed algorithm.  The methodology encompasses an Internet browser extension to autonomously and independently collect information on  the posts presented to users by the News Feed algorithm (Section \ref{sec:measurementmetrics}). Such  information  is not available through the   Facebook API. 

\textbf{Empirical findings} are reported using data collected from a measurement campaign conducted during the 2018 Italian elections.
%
We observed that $a)$ the 
filtering algorithm tends to select information that is aligned with user’s perceived political orientation, $b)$ this effect is more prominent at the topmost News Feed position and $c)$ neutral users are also exposed to non-uniform filtering (Section~\ref{sec:italia}). 

\textbf{An analytical model } is proposed to quantify the visibility and occupancy of publishers in the users News Feeds. The model allows us to conduct a what-if   analysis, to assess the metrics of interest under different filtering mechanisms and is validated using data from the Italian election experiment (Section~\ref{sec:model}).

\textbf{A model-based bias assessment } is conducted using  the Italian dataset.  The dataset is used   to parameterize the proposed model, and yields a baseline publisher visibility (i.e., without the influence of the News Feed algorithm).   The  measured visibility is then contrasted against the baseline to quantify the \emph{bias}, i.e., how publishers' occupancies are affected by user's orientations as they are perceived by the News Feed algorithm (Section~\ref{sec:whatif}).

\section{Measurement Methodology}

\label{sec:measurementmetrics}

The goal of our experiments is to assess the bias experienced by OSN users through a reproducible method. 
To this aim, we  created controlled virtual users that have no social ties and that follow  a preselected set of sources.   By considering minimalistic user profiles, we can assess how preferences towards sources affect  posts presented to users.

In what follows, we first introduce some basic terminology, followed by our data collection methodology, contrasting it against the state of the art.  Finally, we present the metrics of interest which are evaluated in the remainder of this work.

\subsection{Terminology}

\emph{Publisher} is a source that publishes  posts. Each publisher is associated to a public page, which contains all the posts published by that source.  

\emph{News Feed} is the personal page of a given Facebook user, which contains all the posts that Facebook suggests to that user, in an ordered list. We use the terms News Feed and timeline interchangeably.  With some abuse of notation, News Feed also refers to   the algorithm used by Facebook to recommend posts to users.  The distinction between the ordered lists of posts and the algorithm used to fill such lists should be clear from the context.

A user who \emph{follows} a publisher    subscribes to that publisher to receive its posts. A user must follow a publisher  to have post from the publisher's page in the user's News Feed. In our work, all bots follow  the same  set of preselected publishers.

A user who \emph{likes} a  page from a publisher automatically follows that  page. A user likes a page to show general support for that page and its posts. In addition, a user can also like individual posts. In our work,  users orientations are established by letting    them \emph{like}  a subset of the preselected publishers, as well as posts from this subset of publishers.   

\subsection{Data collection methodology}

Next, we present our measurement methodology. 


\textbf{Select representative pages}  
 We  preselect a set of representative  public  Facebook pages (publishers). This selection is subjective in nature, and must account for the classification of publishers into  categories.  In the case of political elections, representative pages include newspapers and the pages of the political parties.

\textbf{Create representative virtual users} 
We create representative  virtual Facebook 
users which capture tendencies on the selected representative pages. Such users are also referred to as \emph{bots}. Each bot follows \emph{all} the  preselected representative pages. In addition, each bot may have an \emph{orientation}, which may be captured by one of a number of mechanisms provided by Facebook. For instance, a bot may 1) ``like'' a page, 2) ``like'' a post from a page  or 3) ``comment'' posts from  a page. We also consider a bot that  does not have an orientation. We refer to it as  \emph{undecided}.

\textbf{Collect  snapshots and impressions} 
Each  bot   kept  open an Internet   browser window  (Firefox or Chrome)  accessing the Facebook page.  The bots were instrumented to collect data on the posts to which they were exposed.  
To that aim, a browser extension,  named Facebook Tracking Exposed, or simply \emph{fbtrex}~\cite{fbtrex} was developed.
The extension auto-scrolls the Facebook window at pre-established instants of the day.  Every auto-scroll produces a  set of posts which are stored at a local database.  Each set of posts is referred to as a \emph{snapshot}.  
Typically, each bot is scheduled to collect thirteen snapshots per day.
  Each post appearing    in a snapshot counts as  a post  \emph{impression}.   
At each bot, Facebook Tracking Exposed collects all impressions and records their corresponding publisher,  publication  time, impression time, content, number of ``likes'' and number of shares.

\textbf{Collect  all  published posts} 
We also have a second dataset which contains the set of \emph{all} posts published by the representative pages during the interval of interest, as provided by the Facebook API.  This dataset is  used to study what  users would experience in the absence of  filters, or in the presence of alternative filters.
%
%
%
%
%
%

\textbf{Compute metrics of interest and analyze  biases}  Bias is defined with respect to a baseline metric, which subsumes a model.  A simple model is introduced in Section~\ref{sec:model} and the corresponding bias is reported in Section~\ref{sec:whatif}.     


\textbf{Share datasets and scripts} By sharing the datasets and scripts, we allow other researchers to reproduce the obtained results and to collect new data. We envision that the collective sampling and analysis of the News Feed is a key step towards building awareness and transparency.

\subsection{State of the art}
Information about post impressions used to be available  in a deprecated version of the Facebook API from 2015.   In any case, that information was not necessarily reliable as recognized by  Facebook itself~\cite{Facebook_API_home}. 
For such reasons, we believe that \emph{fbtrex} and the methodology described in this section constitute  important building blocks to promote transparency  in the Facebook ecosystem. Related work is further discussed in Section~\ref{sec:related}. 

\subsection{Metrics of interest}
Next, we define our key metrics of interest that will be obtained from the dataset generated by the experiment. We consider the top $K$ positions of the News Feed of each user.  
Let $Q_{ij}$ be the number of unique posts from publisher $j$ viewed at bot~$i$.  Let $S_i$ be the number of snapshots taken by bot $i$. 
Then, the measured effective arrival rate of publisher $j$ at bot $i$ is given by  
\begin{equation}
\tilde{\lambda}\pubj=\frac{Q\pubj }{ S_i}.
\label{eq:mes_lambda}
\end{equation}
Let $I_{ij}$ be the number of impressions of publisher $j$ at bot $i$. Let $N_{ij}$ be the average number of posts of publisher $j$ in the top $K$ positions of the News Feed of bot $i$.  $N\pubj$ is given by
\begin{equation}
N\pubj=\frac{I\pubj }{ S_i}.
\label{eq:mes_occupancy}
\end{equation}
Let $\pi_{ij}$ be the fraction   
of snapshots that contain at least one post from publisher $j$ in the top $K$ positions of the News Feed of user $i$.   
We refer to $\pi_{ij}$ and $N_{ij}$ as the \emph{visibility} and the \emph{occupancy}  of publisher $j$ at News Feed $i$, respectively.  The \emph{normalized occupancy}
 is given by $N_{ij}/K$.  The visibility and the normalized occupancy are two metrics of exposure of publishers. They can be interpreted as the probabilities that user $i$ sees a post of publisher $j$  if the user goes  through all the posts in the News Feed or if he/she picks randomly a single post in the News Feed, respectively.  In this paper, due to space limitations we  focus primarily  on the occupancy.


\begin{figure*}[t]
\begin{center}
   \includegraphics[width=0.8\textwidth]{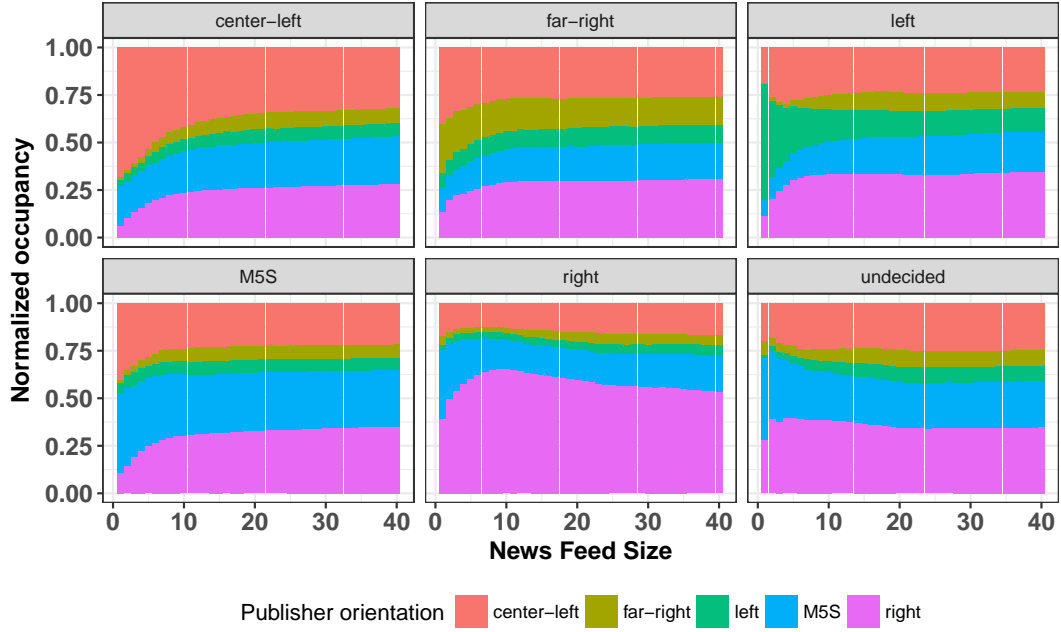}
\end{center}
\caption{Normalized occupancy versus News Feed Size ($K$) at the different bots.}
\label{fig:occupancy_over_k}
\end{figure*}

\section{Experimental Findings} 

Next, we report empirical findings on the 2018 Italian  elections.

\label{sec:italia}

\subsection{Experimental setup}

The Italian  election was held on March 4th  2018, and our experiment was conducted between January 10, 2018 and March 6, 2018, encompassing the preparation for  the election campaign and the reactions to its outcome.

 We asked some Italian voters to select a set of thirty representative  public  Facebook pages, six for each of the following five political orientations: center-left, far-right, left, five-star movement (M5S) and right. Due to space limitations, the selection of representative  pages is reported  in our technical report.\footnote{http://bit.ly/2KfT3Hn} The classification of publishers into political categories may obviously be questionable, but our focus in this paper is on the methodology rather than on specific political conclusions. Moreover,  most of our results are detailed on a per-publisher basis.

Then, we created six  virtual Facebook 
users. Recall that each bot followed \emph{all} the thirty preselected pages. We gave to five bots a specific political orientation, by making each of them like pages from the corresponding publishers. The sixth bot is  \emph{undecided}, and does not like any page.

Each bot was scheduled to collect thirteen snapshots per day. Snapshots were collected once every hour, 
from 7 am to 7 pm (Italian local time).  The collected snapshots and impressions, together with the set of all posts published by the thirty representative pages, obtained through Facebook API, constitute our dataset.

\begin{figure*}[h!]
        \includegraphics[width=0.9\textwidth]     {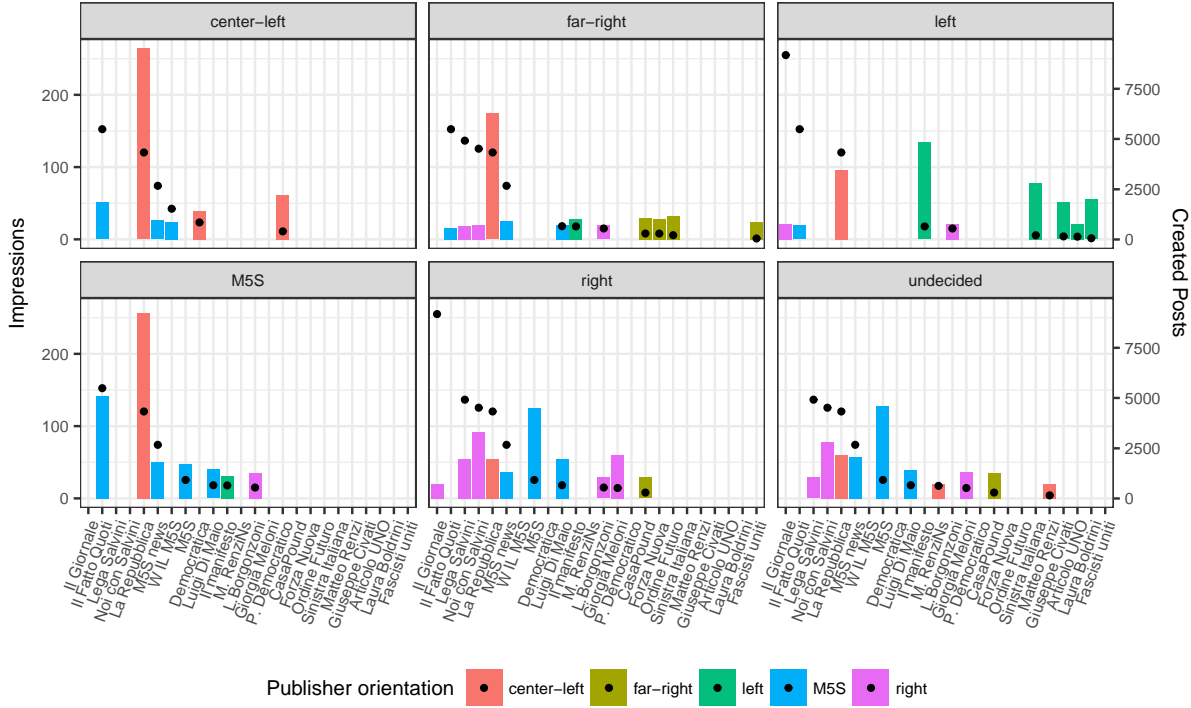} 
\vspace{-0.2in}
\caption{Publishers impressions at the six bots  (bars colored by orientation) and number of created posts (black dots).}
\vspace{-0.1in}
\label{fig:pubfontes}
\end{figure*}

\subsection{Experimental findings}

\paragraph{The effect of filtering is stronger at the topmost News Feed position} \label{sec:topstrong} Figure~\ref{fig:occupancy_over_k} shows the normalized publisher occupancy  as a function of the timeline size. The publishers are ordered and colored according to their political orientation. It can be seen that the occupancy concentration  is higher at the topmost positions, and asymptotic values are reached when  $ K\approx 10$. The noteworthy  bias on the topmost position is referred to in the literature as \emph{ranking bias}~\cite{Search} and must be placed under scrutiny, as  there is a strong correlation between position and click rates~\cite{Bakshy2015,Epstein2015} .

\paragraph{Occupancy reflects orientation}  Figure~\ref{fig:occupancy_over_k}
also shows that the occupancies  reflect the orientation of the bots. For instance, the News Feed of the bot with a center-left orientation was occupied mostly by center-left (red) publishers.  As a notable exception, center-left posts were prevalent in the News Feed of the bot with a far-right orientation. Nonetheless,  the occupancy of  far-right posts in that bot was still the highest among all bots.

\paragraph{Noticeable publishers selection} The bars in Figure~\ref{fig:pubfontes} show the total number of impressions per publisher in the topmost position of the News Feed of each bot (the color of the bars indicates the orientation). For the sake of readability, only publishers that achieved a normalized occupancy larger than $5\%$ are represented in this figure.  The black dots correspond to  the number of posts created by each publisher  
(the publishers are  ordered by the number of posts created).  Figure~\ref{fig:pubfontes} shows that only a small subset of publishers are represented in  topmost positions. For example, the center-left bot sees only posts from center-left and M5S publishers. Moreover, the number of impressions per publisher is not proportional to the number of posts the publisher created, a further indication of a filtering effect from News Feed algorithm.  


\paragraph{Neutral users are also exposed to non-uniform filtering} 
It is worth noting that filtering affects also the “undecided” bot, with some publishers over-represented in the News feed.
\vspace{-0.05in}
\section{ News Feed Model}
\label{sec:model}



\paragraph{Model description} Next, we derive an analytical model for a FIFO implementation of the News Feed. 
We assume that a News Feed has $K$ slots, new posts are inserted at the top of the News Feed and each new arrival shifts older posts one position lower.
A post is evicted from the News Feed when it is shifted from position $K$.

%
Let $\mathcal{I}$ be 
the set of $I$ users, and let  $\mathcal{J}$ 
be the set of $J$  publishers: $\mathcal{J}_i$ denotes the set of publishers 
followed by user $i \in \mathcal{I}$.  
Publisher $j \in \mathcal{J}$  publishes posts according to a Poisson process with rate $\Lambda_j$. The total publishing rate is $\Lambda=\sum_{j=1}^J \Lambda_j$.   Let $\lambda\pubj \leq \Lambda_j$ be the effective arrival rate of posts published by $j$ in the News Feed of user $i$. We denote by $\lambda_i$  the aggregate rate of  posts published in the News Feed of user $i$, $\lambda_i=\sum_{j=1}^J \lambda_{ij}$. We further let  $\lambda_{i,-j}$ be the arrival rate of posts in the News Feed of user $i$ from all publishers other than  $j$, $\lambda_{i,-j}= \lambda_i-\lambda_{ij}$. All the variables used are summarized in Table \ref{tab:notation}.


\begin{table}[h]
\centering
\begin{tabular}{l l}
\hline
\hline
Variable & description \\
\hline
$j$ & $j$-th publisher \\
$i$ & $i$-th News Feed user \\
$\Lambda_j$ & post creation rate by publisher $j$\\
$\lambda_{ij}$ & arrival rate of posts from $j$ at user $i$ \\
$\lambda_i$ & total arrival  rate of posts at user $i$\\
\hline
\hline
\multicolumn{2}{c}{Metrics of interest as estimated by the model} \\
\hline
$h_{ij}^{(m)}$ & hit probability of publisher $j$ at user $i$\\
$\pi_{ij}^{(m)}$ & visibility of publisher $j$ at user $i$ \\
$N_{ij}^{(m)}$ &  occupancy of publisher $j$ at user $i$ \\
\hline
\multicolumn{2}{c}{Metrics of interest as obtained from measurements} \\
\hline
$\pi_{ij}$ & measured visibility of $j$ at $i$\\
$N_{ij}$ & measured occupancy  of $j$ at $i$\\

\hline
\hline
\end{tabular}
\caption{Table of notation}
\label{tab:notation}
\end{table}

\paragraph{Metrics of interest} We consider the News Feed of user $i$.
Subscript $(m)$ is used  to denote metrics computed using the analytical model.
The visibility of publisher $j$ is given by 
$\pi_{ij}^{(m)}= 1-({\lambda_{i,-j}}/{\lambda_i})^K$, 
and    the rationale goes as follows. After every new arrival,  with probability ${\lambda_{i,-j}}/{\lambda_i}$ the topmost post of publisher $j$ will be shifted down by one unit.  After $K$ consecutive shifts, which occur with probability $({\lambda_{i,-j}}/{\lambda_i})^K$, publisher $j$ will not be visible at the News Feed of user $i$. The occupancy of contents of publisher $j$, in turn, follows from Little's law and  is given by
\begin{equation}\label{eq:occupancy_fifo}
N_{ij}^{(m)}= {\lambda\pubj K}/{\lambda_i}. 
\end{equation}
When $K=1$ we have $N_{ij}^{(m)}=\pi_{ij}^{(m)}$. As the filtering effect is stronger on the topmost  position (Section~\ref{sec:italia}), except otherwise noted, we let $K=1$.

\begin{figure}[h!]
\begin{center}
  \includegraphics[width=0.5\textwidth,height=2.5in]{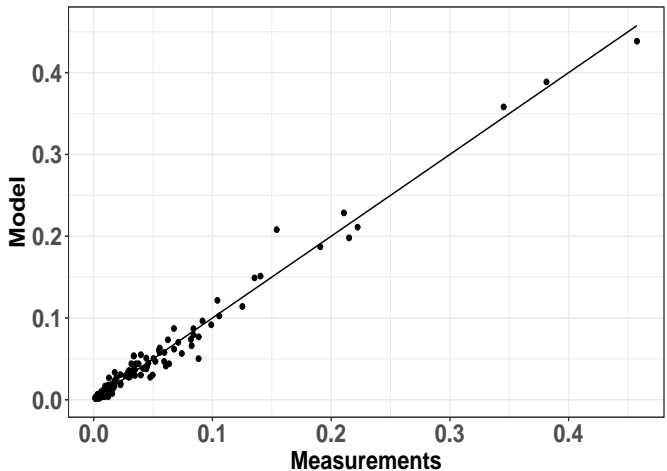}
\end{center}
\vspace{-0.1in}
\caption{Model-predicted against measured occupancies.}
\label{fig:valid}
\vspace{-0.1in}
\end{figure}

\paragraph{Model  validation}
A preliminary model validation using data from the 2018 Italian elections is now introduced.
 In Figure~\ref{fig:valid}, each point  corresponds to a publisher-user pair.  A point $(x=N_{ij},y=N_{ij}^{(m)})$ indicates that, for the given pair, 
an occupancy   $N_{ij}^{(m)}$ estimated by the proposed model using eqs.~\eqref{eq:occupancy_fifo} and~\eqref{eq:mes_lambda}  corresponds to a  measured occupancy $N_{ij}$. Most of the points are close to the $N_{ij}=N_{ij}^{(m)}$ line, indicating the expressive power of the model.  Note that to generate Figure~\ref{fig:valid} we relied on the effective arrival rates of posts, $\lambda_{ij}$.  Future work consists of inferring such rates directly from publishers and users profiles, e.g., using machine learning techniques.   

\begin{figure*}[t]
\begin{center}
   \includegraphics[width=0.95\textwidth]{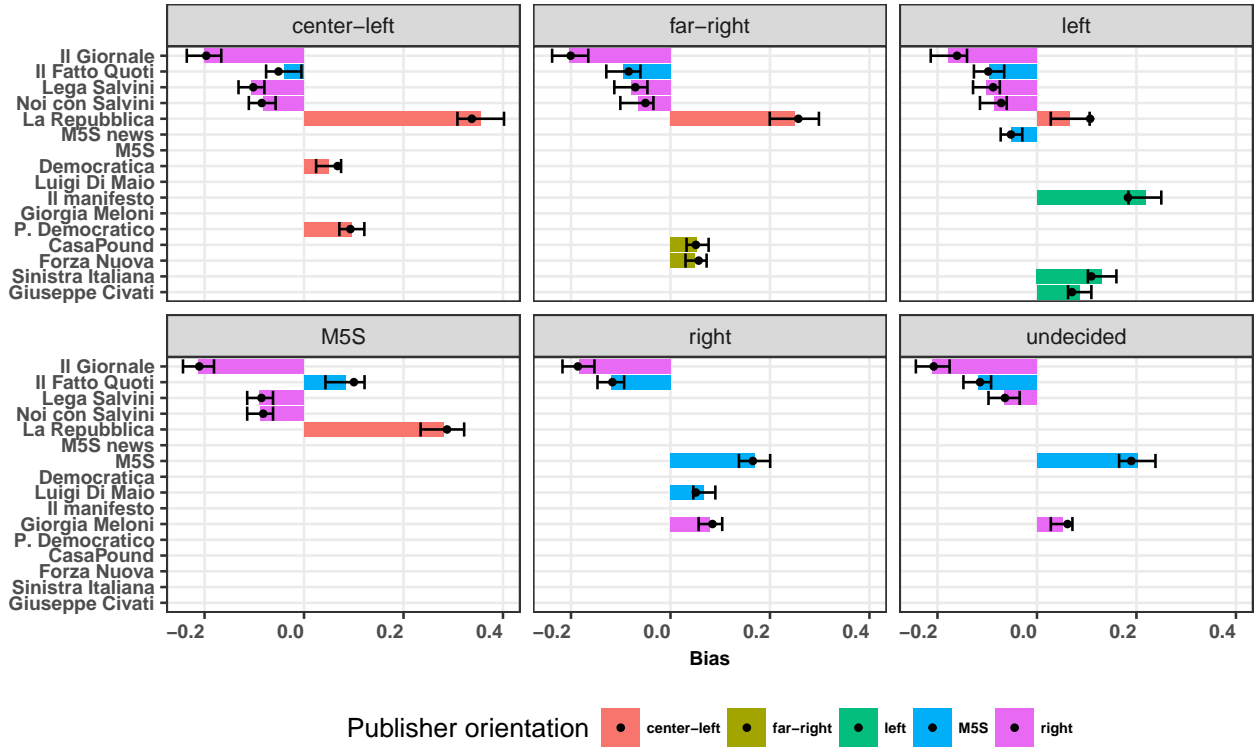}
\end{center}
\vspace{-0.1in}
\caption{Bias per top publishers at each bot. }
\label{fig:contrafa1}
\vspace{-0.25in}
\end{figure*}

\section{  Quantifying Bias: A Model-Based Approach} \label{sec:whatif}

We provide further insight on the influence of the News Feed personalization algorithm. To this aim,  we propose a model-based bias analysis.
%
%

Recall that $N_{ij}^{(m)}$ is the occupancy computed from \eqref{eq:occupancy_fifo}, using the effective arrival rates $\lambda_{ij}$ of publisher $j$ at bot $i$.
Alternatively, let $N_{ij}^{(u)}$ be the occupancy of publisher $j$ at bot $i$ in the \emph{unfiltered} scenario wherein all posts published appear in the feed.  
$N_{ij}^{(u)}$ is obtained from~\eqref{eq:occupancy_fifo}, letting $\lambda_{ij}=\Lambda_j$ and $\lambda_i =\sum_j \Lambda_j$.  
%
%
We observe that $N_{ij}^{(u)}$ is the same for all the users and we denote it simply as $N_{j}^{(u)}$.


We define the bias as the difference between the two occupancies, i.e. 
\begin{equation} \label{eq:bias1}
b_{ij}=N^{(m)}_{ij} - {N}^{(u)}_{j}.
\end{equation}
Note that the definition of bias is general, and can be coupled with different base-line models of occupancy.  Utility-driven models of visibility, inspired by~\cite{Dehghan2016a,netecon,Kelly1997}, are currently one of our subjects of study.  

Figure~\ref{fig:contrafa1} shows  bias estimates obtained from the Italian elections experiment.
Black dots indicate  model predictions of bias, obtained directly from the measurements together with equation~\eqref{eq:bias1}. 
We  evaluated $95$\% confidence intervals  for the bias using a bootstrapping model-free non-parametric resampling method. To this purpose, one thousand  virtual experiments have been conducted.  Each experiment consisted in sampling, with replacement, from the set of posts collected by \emph{fbtrex} (to estimate confidence intervals for $N_{ij}$) and from the set of posts obtained from the Facebook API dataset (to estimate confidence intervals for $N^{(u)}_{ij}$). Bars and error bars 
show the average points and the boundaries of the 95\% confidence intervals for the bias.

Figure~\ref{fig:contrafa1} provides further insights on the findings reported in Section~ \ref{sec:italia}.  In particular, we observe a strong positive bias towards La Republica at the far-right and M5S bots and a strong negative bias towards Il Giornale at all bots. In addition, the  bias experienced by the left-oriented user is
particularly  well aligned with its profile. Note also that there is a
strong positive bias towards M5S posts at the undecided bot. 
M5S was the party with the largest number of votes in the 2018  Italian general election. 
We also observed that the bias profile of the undecided neutral user is similar to that of the right-oriented one.


\section{Related Work}


\label{sec:related}

There is a vast literature on how the Facebook News Feed works, including topological aspects related to cascading structures and growth ~\cite{Gjoka2010,anatomy_fB,Cheng2018} and its effects on the creation of echo chambers and  polarization~\cite{BESSI2016319,bessi2016users}. 
In this paper, we  study the News Feed filtering impact on the dissemination of information,  measuring and modeling the visibility of publishers and posts in the News Feed.

\subsection{Public datasets and reproducible methodologies}

The behavior  of users searching for visibility was studied in    
~\cite{eslami2016first,Bucher2012,Sleeper2013}. Such studies are primarily based on small datasets.  A notable exception is~\cite{Bakshy2015,Bakshy2012}, who   considered a massive dataset provided by Facebook through restrictive non-disclosure agreements.  Datasets   to assess Facebook publishers'  visibilities are  usually not made publicly  available. Our work aims to contribute by filling that gap.

The proposed methodology  meets the set of principles intended to ensure fairness in the evolving policy and technology ecosystem introduced by the ACM  \cite{Garfinkel2017}: awareness, access and redness, accountability, explanation, data provenance, auditability, and validation and testing.  We particularly focus on  awareness, explanation and auditability, as we do not rely on the Facebook API to collect impressions.

In this paper we  profile Facebook, which is taken as a black box to be   scientifically analyzed. 
This approach dates back to Skinner tests~\cite{skinner_box}, and   
 has been gaining significant attention in the literature of social networks~\cite{Tan2017}. 

\subsection{Bias and visibility models}

The awareness of users about the News Feed algorithm was evaluated in~\cite{Eslami2015a}. Eslami et al. reported that $62,5\%$ of the interviewed Facebook users were unaware of the News Feed algorithm. Users believed that every post from their friends and followed pages appeared in their News Feed.

The bias in algorithms, and forms to audit it, were investigated in~\cite{Diakopoulos2013,Sandvig2014}. In~\cite{Epstein2015} it was shown that search engine ranking manipulation can influence the vote of undecided citizens. This bias was further quantified in \cite{Kulshrestha2017}. With the help of our model, we can quantify the bias in the Facebook News Feed.


News Feed models were previously proposed in~\cite{altman2013competition,dhounchak2017viral} and~\cite{netecon}. In this paper, we propose a simple News Feed model and validate it  using real-world data. It is out of the scope of this paper to present a nuts-and-bolts perspective on how the Facebook News Feed works. Instead, our goal was to provide a simple model that can  explain the occupancy and visibility of different publishers, given a reproducible measurement framework. As in any scientific experiment, we  cope with a tradeoff between model expressive power and simplicity.  In this preliminary work, we considered the simplest possible News Feed model, namely a FIFO model, that produces meaningful results.

The implications  of the limited budget of attention of users in OSNs have been previously studied by Reiffers-Masson \emph{et al.}~\cite{masson2017posting} and  Jiang \emph{et al.}~\cite{jiang2013optimally}.  In these two papers, the authors consider the problem of optimally   deciding what  to post and to  read, 
 respectively.  Such works are complementary to ours. To the best of our knowledge, none of the previous works considered the problem of inferring the visibility of publishers from News Feed measurements.

\section{Discussion}

\textbf{FIFO model} The proposed FIFO model is a preliminary step to capture  News Feed dynamics. The stronger filtering at the topmost News Feed positions,  that we observed in Section~\ref{sec:italia}, cannot be justified by FIFO operation.   Nonetheless,  this simple model already allows us to appreciate what kind of analysis can be conducted  by combining measurements and analytical tools. We are currently working on a more complex model, e.g., to account for different residence times for different posts using time-to-live counters~\cite{Dehghan2016a,Kelly1997,netecon}.  

\textbf{Small number of  users} Our experimental methodology encompasses a small number of bots (users), as each bot must go through a full Facebook account registration. In particular, in our case study we considered six bots.   We believe that such bots already provide a representative perspective on the biases introduced by Facebook, and we use bootstrapping techniques to generate  additional samples from our observations. 
Future work consists of increasing the number of bots to tens of bots, including bots with similar profiles to compare their impressions. 

\textbf{Compliance to Facebook policies} Our methodology encompasses the creation of fake users, as we  use a limited number of experimental Facebook accounts to follow and like public pages.  Such accounts have a very low impact on the system, as they do not have real or virtual friends. To minimize potential harm, we remove our experimental Facebook accounts after we are done with our experiments.

\section{Conclusion}
\label{sec:concl}

In this work, we proposed a framework for reproducible measurements and analysis of  the Facebook News Feed algorithm. 
The framework encompasses an  analytical model  which enables quantitative what-if analysis to assess the bias introduced by the News Feed algorithm.

%
%
%
%
%
We were able to conclude that the algorithm tends to reinforce the orientation indicated by users about the pages they ``like'', by filtering  posts and creating biases among the set of followed publishers. The effects of filtering are stronger on the topmost position where only a fraction of the set of publishers followed  by the users was represented. We observed that a neutral user that did not ``like'' any page was also exposed to a noticeable bias.

Facebook mission  is to ``give people the power to build community.'' We believe that the measurements, model and tools presented in this work are one step further towards that goal, as they help evaluating algorithms' transparency and promote user awareness about the filtering process they are submitted to.   Ultimately, such awareness is key to protect and empower Facebook users,  communities, society and democracy as a whole.  

\section*{Acknowledgements} 
This work was partially funded by the Brazilian-French project THANES (co-sponsored by FAPERJ and INRIA). 
D.~Menasché was also partly sponsored by a JCNE fellowship (FAPERJ), CNPq and CAPES. 

\begin{footnotesize}

\bibliographystyle{plain}
\bibliography{references}

\end{footnotesize}

\end{document}